\documentclass[conference]{IEEEtran}
\IEEEoverridecommandlockouts
\usepackage{url}
\usepackage{cite}
\usepackage{amsmath,amssymb,amsfonts}
\usepackage{algorithmic}
\usepackage{graphicx}
\usepackage{textcomp}
\usepackage{xcolor}
\usepackage{siunitx}
\usepackage{booktabs}
\def\BibTeX{{\rm B\kern-.05em{\sc i\kern-.025em b}\kern-.08em
    T\kern-.1667em\lower.7ex\hbox{E}\kern-.125emX}}
\usepackage[export]{adjustbox}
\usepackage{flushend}

\newcommand{\linebreakand}{%
  \end{@IEEEauthorhalign}
  \hfill\mbox{}\par
  \mbox{}\hfill\begin{@IEEEauthorhalign}
}

\begin{document}



\title{Google Search Advertising after Dobbs v.~Jackson}


\author{\IEEEauthorblockN{Yelena Mejova}
\IEEEauthorblockA{
\textit{ISI Foundation}\\
Turin, Italy \\
yelenamejova@acm.org}
\and
\IEEEauthorblockN{Ronald E. Robertson}
\IEEEauthorblockA{\textit{Cyber Policy Center} \\
\textit{Stanford University}\\
Stanford, CA, USA \\
rer@acm.org}
\and
\IEEEauthorblockN{Catherine A.~Gimbrone}
\IEEEauthorblockA{\textit{Department of Epidemiology} \\
\textit{Columbia University}\\
New York, NY, USA \\
ag2246@cumc.columbia.edu}
\and
\IEEEauthorblockN{Sarah McKetta}
\IEEEauthorblockA{\textit{Department of Population Medicine} \\
\textit{Harvard Medical School and }\\
\textit{Harvard Pilgrim Healthcare Institute}\\
Boston, MA, USA \\
sarahmcketta@fas.harvard.edu}
}




\maketitle
\IEEEpeerreviewmaketitle

\begin{abstract}

Search engines have become the gateway to information, products, and services, including those concerning healthcare. Access to reproductive health has been especially complicated in the wake of the 2022 Dobbs v. Jackson decision by the Supreme Court of the United States, splintering abortion regulations among the states. In this study, we performed an audit of the advertisements shown to Google Search users seeking information about abortion across the United States during the year following the Dobbs decision. We found that Crisis Pregnancy Centers (CPCs)--organizations that target women with unexpected or ``crisis'' pregnancies, but do not provide abortions--accounted for 47\% of advertisements, whereas abortion clinics -- for 30\%.  Advertisements from CPCs were often returned for queries concerning information and safety. The type of advertisements returned, however, varied widely within each state, with Arizona returning the most advertisements from abortion clinics and other pro-choice organizations, and Minnesota the least. The proportion of pro-choice vs. anti-choice advertisements returned also varied over time, but estimates from Staggered Augmented Synthetic Control Methods did not indicate that changes in advertisement results were attributable to changes in state abortion laws. Our findings raise questions about the access to accurate medical information across the U.S. and point to a need for further examination of search engine advertisement policies and geographical bias.

\end{abstract}


\section{Introduction}
\label{sec:intro}

In June 2022, The United States Supreme Court overturned a federal right to abortion in its rule for Dobbs v. Jackson\cite{Dobbs2022}.
As a result, state-level abortion laws, which already varied widely prior to the court decision \cite{llamas2018public},determine who can get an abortion, when, and under what circumstances\cite{davis2022state}.
In the wake of Dobbs v. Jackson, abortion became illegal under any circumstances (i.e., a total ban) in 13 states\cite{nytabortion}. 
In 4 states, restrictions were placed on procedures as early as 6 weeks gestation, amounting to a de facto ban, since the majority of pregnancies are not detected until 6 weeks or later\cite{watson2022frequency}.

Even prior to Dobbs v. Jackson, people seeking abortions experienced barriers to accessing care.
In 2014, 18\% of people seeking abortion traveled over 50 miles each way to obtain care\cite{fuentes2019distance}, 
and since Dobbs v. Jackson, the median travel distance has increased to 191 miles\cite{mcguinn2024changes}. 
In 2023, nearly 1 in 5 abortion patients traveled out of state to access abortion care\cite{sheets2023high}.
As a result of these barriers, an increasing proportion of abortion care has moved to telehealth and online delivery of abortion medication\cite{koenig2023role,brown2023provision} as well as self-managed abortion.
Therefore, search engines have become critical tools for people seeking abortion--not only to identify information regarding local laws and local abortion services, but also to identify telehealth services and information on self-managed abortion\cite{flores2023internet, Mejova_Gracyk_Robertson_2022, jerman2018people}. 
In the wake of the Dobbs v. Jackson decision, searches related to abortion and contraception increased, particularly in states with laws limiting abortion access\cite{gupta2023trends}.

Crisis pregnancy centers, or CPCs, are organizations that have played a large patient-facing role in the anti-choice movement for years. 
These businesses target patients seeking abortion, and then often provide misinformation to discourage abortion procedures\cite{thomsen2023presence}. 
A core part of their strategy is positioning or advertising themselves as abortion providers\cite{arthur2023examining}. 
In the past decade, CPCs have contributed to the "infodemic"\cite{pagoto2023next} of online misinformation about abortion efficacy, safety, and availability--in part due to search engine optimization and online advertising, such that CPCs are heavily represented in Google searches for terms related to abortion\cite{pagoto2023next, Mejova_Gracyk_Robertson_2022}.

\begin{figure}[t]
\centering
\includegraphics[width=0.95\columnwidth]{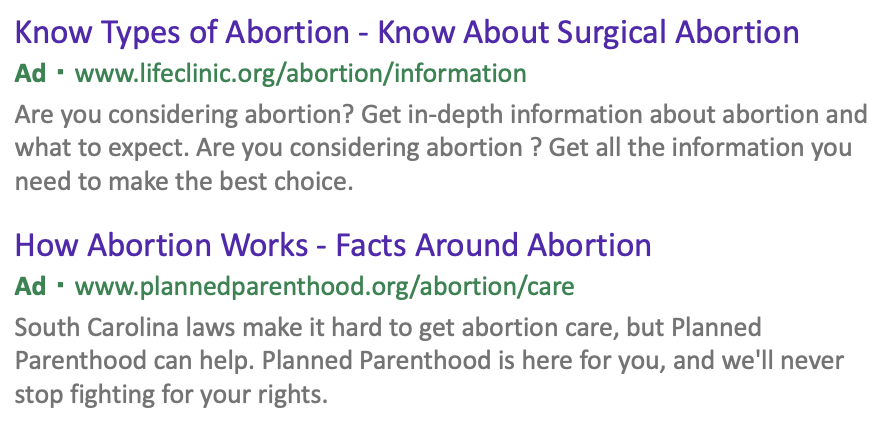}
\caption{\textbf{Example of two advertisements on Google Search.} }
\label{fig:example}
\end{figure}

Advertising is an important source of revenue for Google (for an example, see Figure \ref{fig:example}), and paid Google advertisements (a.k.a. ads) are a cost-effective form of marketing for businesses\cite{almestarihi2024measuring}. Buying advertising on Google is a strategy employed by businesses and websites promulgating misinformation and fake news\cite{papadogiannakis2023funds}, including CPCs\cite{swartz2021comparing}. 
Google allows advertisers to tailor their audience, such that advertisements vary widely based both on search terms and the location of the user searching\cite{ahmadi2024overwhelming}.
In the context of people seeking abortion, Google advertisements may play a large role in paradoxically directing towards legitimate services and misdirecting towards misinformation. 
While several states ban abortion clinics from advertising their services, these bans do not impact online advertisements (i.e., via Google or other search engines)\cite{lawatlas}.

Several studies documented an increase in Google searches related to sexual health and abortion access resulting from the Dobbs decision. 
For example, analyses of Google search trends demonstrated between a 25-62\% increase in searches using terms related to abortion or abortion medication after the Dobbs decision was made public \cite{poliak2022internet, gupta2023trends}.
This increase was patterned by state abortion policy--as states became more restrictive, the number of searches increased in an apparent dose-response manner \cite{poliak2022internet}.
Searches for vasectomy also spiked in an apparent dose-response manner \cite{patel2022search,sellke2022unprecedented}.

Not only did people respond to the Dobbs decision by using Google to search for information related to abortion and sexual health, but the results and content that Google made available to users also adapted to the changing policy environment. 
Google consistently updates its algorithm to tailor its search results; in fact, in 2019, in response to public pressure, Google updated its advertising policy to require that all advertisers disclose whether they offered abortion services (to distinguish abortion providers from CPCs) \cite{dodge2018quality, dodge2022influence}.
After this change, advertisements for abortion services were more likely to facilitate access to abortion, though a third of advertisements still led users to businesses that would hinder abortion access \cite{dodge2022influence}. 

In our prior work examining how the content of Google search results varies by state, we used 10 abortion-related search queries to conduct Google searches from 467 different locations in the United States and examined the distribution of abortion clinics and CPCs resulting from those searches \cite{Mejova_Gracyk_Robertson_2022}. 
We found that locations with a higher proportion of abortion clinics yielded more search results for abortion clinics, and that locations with fewer abortion clinics yielded more results for CPCs. Users searching for abortion content were less likely to encounter a result including an abortion clinic if they were searching from a state with restrictive legislation. Overall, the searches yielded relatively accurate information and directed users to abortion clinics 79\% of the time, and CPCs only 7\% of the time. Critically, this research did not examine the content of advertisements, nor did it include searches after the Dobbs v. Jackson case.

While prior research has demonstrated the use of Google searches to identify information related to abortion services,\cite{gupta2023trends, flores2023internet} as well as the proliferation of CPC website promoting misinformation in Google results, \cite{pagoto2023next, Mejova_Gracyk_Robertson_2022} no research to date has examined the content of Google advertisements that appear in searches related to abortions, nor to what extent this advertising content varies by US state or state abortion policy climate.

In the present study, we use best practices in algorithm auditing~\cite{metaxa2021auditing} to examine the content of advertisements that would have been shown to people searching for abortion-related content on Google after Dobbs v. Jackson.
%
%
Our study follows prior work by conducting a focused set of abortion-related queries from a wide array of locations across the US~\cite{Mejova_Gracyk_Robertson_2022}, and advances on that work by collecting a more temporally dense dataset of searches (daily instead of weekly) for a year following Dobbs v. Jackson.
Given the dynamic nature of the web and the constant updates that Google Search undergoes, our dataset captures a distinct set of search results from prior work, and does so in the wake of a historic event impacting abortion care in the US.



\section{Data Collection and Methods}
\label{sec:data}

\textbf{Queries.} To collect our search data, we first obtained a set of queries that might be used by people who are looking for abortion-related services or information on Google Search. 
Specifically, we used the set of queries developed in past work \cite{Mejova_Gracyk_Robertson_2022}, which began with the query ``abortion'', and used Google's autocomplete search suggestions, People Also Ask, and Related Searches features to find other possible wordings or related queries.
Our final set of 20 queries can be grouped as follows:

\begin{enumerate}

    \item General: \emph{abortion, abortion clinic, abortion clinic near me, abortion clinic (localized), abortion center, abortion center near me, abortion center (localized), pregnancy termination}

    \item Informational:  \emph{kinds of abortion, types of abortion, how does abortion work}

    \item Safety:  \emph{is abortion dangerous, is abortion safe}

    \item Cost:  \emph{abortion cost, free abortion, does insurance cover abortion}

    \item Legality:  \emph{is abortion illegal in my state, is abortion legal in my state, is abortion illegal, is abortion legal}
    
\end{enumerate} 

The search queries containing ``(localized)'' were localized by adding the name of the location from which the search was going to be conducted. 

\vspace{0.1cm}
\textbf{Locations.} We conducted the searches with our set of queries on a daily schedule using an open source python library called WebSearcher~\cite{robertson2020websearcher}.
This library allowed us to specify the location from which the search was conducted (e.g., a particular US congressional voting district), which allowed us to compare the search results returned for the exact same query from different locations at roughly the same time.
Distinct from prior work, which used 467 counties that were selected via stratified random sampling~\cite{Mejova_Gracyk_Robertson_2022}, here we used the 435 US congressional districts as our set of locations to search from.
After combining our query and location sets, we collected the search results for each query-location pair every day between 2022-06-27 and 2023-06-16.  
In total, we collected \num{1387550} search engine result pages (SERPs) and used the WebSearcher library to parse them and extract the returned advertisements (if any exist on the page). 

\vspace{0.1cm}
\textbf{Advertisements.} We extracted the advertisements from each collected SERP, which contain the results for a particular query, ran on a particular day, localized in a particular district. 
For each ad, we obtained its text and the URL it links to, as well as subtext (if any) and its position on the page (whether it is shown first, second, etc.).
An example of two ads can be seen in Figure \ref{fig:example}.
Note that, because the querying was performed in various locations and over several days, the same ad could be returned in many SERPs. 
Thus, the dataset contains a total of \num{3646559} search result ads, out of which \num{298662} are unique (considering the ad's URL, title, and text snippet). 
We further processed the URLs to extract the top-level domain name (such as \emph{plannedparenthood.org}) in order to aggregate the ads.

\vspace{0.1cm}
\textbf{Domain Annotation.} In order to characterize the organizations linked to these ads, we extracted the domain name (e.g, \emph{plannedparenthood.org}) from each ad's URL and manually annotated them, expanding on domain labels from prior work\cite{Mejova_Gracyk_Robertson_2022}.
Two independent experts specializing in reproductive health policy labeled each domain (K = 0.85, indicating almost perfect agreement) \cite{landis1977measurement}, cross-checking with the Internet Archive in cases of dead links and with a database of crisis pregnancy centers as needed \cite{cpcmap}. 
Conflicts were resolved by consensus between the two coders, and the final set of domain labels was:

\begin{itemize}

\item Abortion clinic (AC)
\item Pro-choice advocacy group or research
\item Pro-choice link to care
\item Crisis Pregnancy Center (CPC)
\item Anti-choice advocacy group or research
\item Anti-choice link to care
\item Other non-profit organization
\item Adoption resource
\item Governmental resource
\item General clinic / medical services
\item News \& blogs
\item Religious organization
\item Dead link

\end{itemize}

\vspace{0.1cm}
\textbf{State Abortion Laws.} In order to contextualize the time period in which we collected our searches, we compiled a list of state-level laws active during the study period. 
To account for the rapidly-changing policy environment, we iteratively tracked these legal developments using a combination of news coverage and databases from policy watchdogs \cite{kff2025,naral2025}, specifically noting the gestation limit (in weeks) until which abortion could be lawfully performed in each state. 
In some states, instead of restrictions based on gestation, the law specified ``fetal viability,'' which can range in interpretation from the late 2nd to 3rd trimester \cite{han2018blurred}. 
Because the definition of fetal viability is generally at the discretion of a medical provider, these laws were considered to be minimally restrictive with respect to gestational limit.  
The laws were categorized as a total ban (most restrictive), 1st trimester (0-12 weeks), 2nd trimester (13-26), or 3rd trimester/fetal viability (27-40 weeks, least restrictive). 
We have made this information available to the research community.\footnote{\url{https://tinyurl.com/AbortionLawsByStateAfterDobbs}}

\vspace{0.1cm}
\textbf{Statistical Methods.}
We employed Staggered Augmented Synthetic Control Methods (SASCM) to estimate the average treatment effect on the treated (ATT) of state-level abortion policy changes on ad content \cite{ben2022synthetic}. 
This method attempts to address causal inference challenges when policies are adopted by different units at different times.
For these analyses, we chose to dichotomize policies as ban (including 6-week gestational limits, given they amount to a de facto abortion ban \cite{watson2022frequency}) or no ban. We chose this more parsimonious approach, rather than the categorical approach to coding policies, both in order to examine the differences across the most extreme restrictions and to provide a sufficient number of both treatment and control states. Ads were dichotomized as proportion pro-choice or anti-choice ad content. 
Both policies and ad content were further aggregated to the weekly level,  with weekly policies coded based on whichever policy was in effect for the majority of days within that week. 
We tested two models based on policy adoption: (1) treatment states that restricted policies from no ban to ban (Idaho, West Virginia) and (2) treatment states that relaxed policies from ban to no ban (South Carolina, Ohio) during the study period. 
Control states were selected so that they matched the pre-study period policies in treatment states and did not change policy categories during the study period. 


\section{Results}
\label{sec:results}



\subsection{Categories of Ads Returned}



\begin{figure}[t]
\centering
\includegraphics[width=0.93\columnwidth]{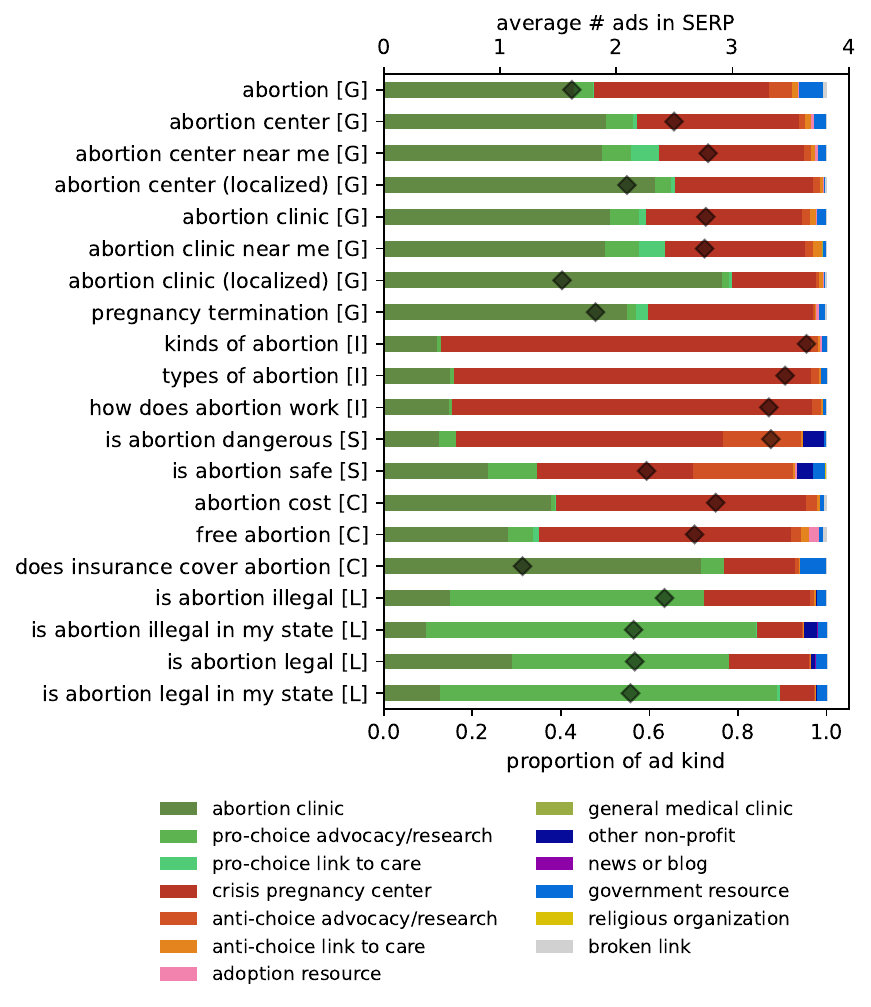}
\caption{Proportion of ad results that are of a particular type (bars; bottom y-axis) and average number of ad results in a SERP (diamonds; top y-axis), by query (type in brackets; General, Informational, Safety, Cost, Legality). }
\label{fig:domain_kind_by_query}
\end{figure}

Overall, the most frequent category of domains in our data was CPCs at 46.8\%, followed by abortion clinics (AC, 29.9\%), pro-choice advocacy or research (16.4\%), and anti-choice advocacy or research (2.4\%).
Aggregating over the study period and across locations, we grouped the results by query type and found substantial variance in the number of ads returned and the proportion of ad types (Figure \ref{fig:domain_kind_by_query}). 
First, the number of ads (the black diamonds, top $x$ axis) ranged from just above 1 (in response to ``does insurance cover abortion'') to around 3.5 (in response to the informational queries). 
Second, queries associated with information seeking (``types/kinds of abortion'', ``how does abortion work'') and its potential dangers (``is abortion dangerous/safe'') returned the largest proportion of CPC and anti-choice ads (colored bars, bottom $x$ axis).
Third, queries about the legality of abortion, returned the largest proportions of pro-choice advocacy ads. 
Fourth, general queries about abortion (including those related to abortion centers or clinics) returned a notable proportion of CPC ads (ranging from: 19\% to 40\%). 


\begin{figure}[t]
\centering
\includegraphics[width=0.85\columnwidth]{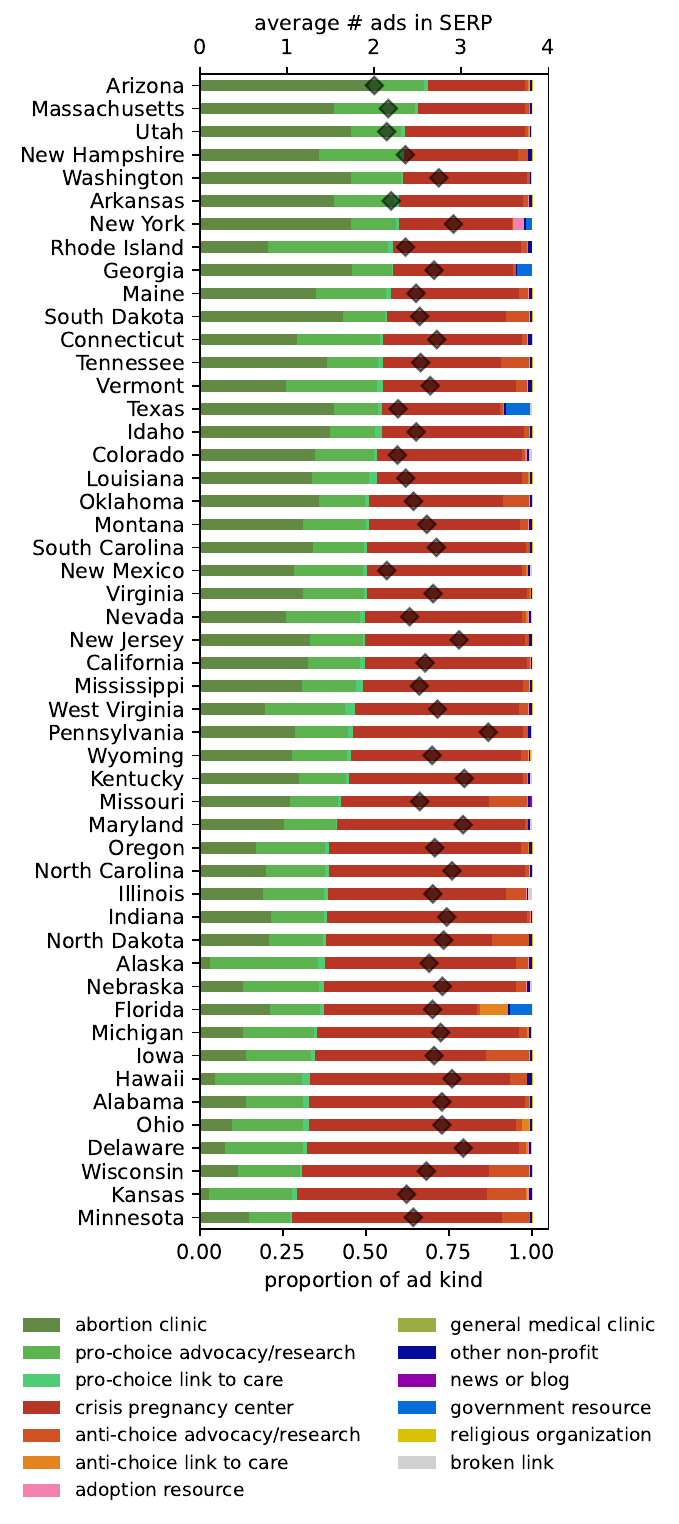}
\caption{Proportion of ad results that are of a particular type (bars; bottom x-axis) and average number of ad results in a SERP (diamonds; top x-axis), by state. The states are ordered by the total proportion of abortion clinic and pro-choice advocacy/research or link to care ads.}
\label{fig:domain_kind_by_state}
\end{figure}

\begin{figure}[t]
\centering
\includegraphics[width=0.99\columnwidth]{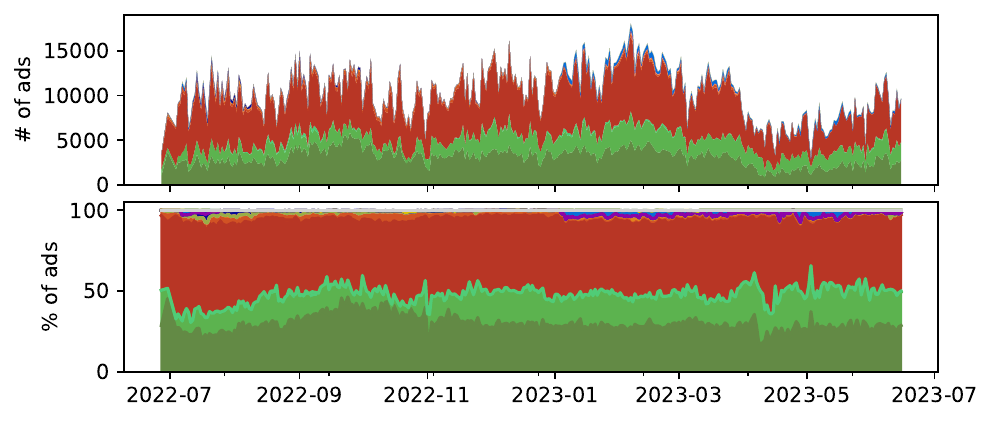}
\caption{Volume of ads returned (top) and percentage of ads (bottom) by kind. Colors are the same as those used in Figure~\ref{fig:domain_kind_by_query}: shades of green correspond to abortion clinics and pro-choice resources while shades of red correspond to CPCs and anti-choice resources.}
\label{fig:domain_kind_by_time}
\end{figure}

Next, we examined SERPs returned for queries executed from each state (aggregated over all queries, times, and locations within each state).
Similar to Figure~\ref{fig:domain_kind_by_query}, Figure \ref{fig:domain_kind_by_state} shows the percentage of each domain type and the average number of ads in a SERP, now grouped by state instead of query type. 
States are sorted by the percentage of ads that were labeled as AC, pro-choice advocacy/research, or pro-choice link to care.
We found that states with the largest proportion of advertising for either ACs or pro-choice resources (and lower proportions of ads for CPCs and anti-choice websites) were Arizona, Massachusetts, Utah, New Hampshire, Washington, and Arkansas.  
Conversely, states with the smallest proportion of such ads (and proportionally more CPCs and anti-choice websites) included Minnesota, Kansas, Wisconsin, followed by Delaware, Ohio, Alabama, and Hawaii. 
Upon visual inspection of these results, no apparent state pattern emerged with regard to notable state features -- i.e., geographical region, population density, or state conservatism, though we did not formally test associations with state characteristics.  

The overall number of ads shown to users in different states varied from approximately 2 to 3.5 per SERP. 
Further, the average number of ads per state was associated with the distribution of ad content per state.
That is, states with more ads generally returned fewer pro-choice resources (Pearson correlation $r$ = -0.49, $p$ = 0.0003) and more anti-choice resources ($r$ = 0.48, $p$ = 0.0004). 


Figure \ref{fig:domain_kind_by_time} shows the type of ads returned each day in our study period by count (top) and as percentages (bottom).
We found wide variation in the total number of ads returned across all searches from one day to another, peaking in February 2023 at over 15,000. 
Although there was some heterogeneity in the proportion of anti-choice ads versus pro-choice ads over time, the balance between the two remained largely similar across time, with ACs and pro-choice resources making up 47.7\% of daily ads (std=0.054), and anti-choice ads comprising 49.9\% (std=0.053).

\subsection{Top Advertisers}


Table \ref{tab:top_domains} shows the top 30 domains, out of a total of 7,969 unique domains across the full dataset, the proportion of ads that led to those domains, and their respective domain type labels. 
Overall, no individual domain dominated the share of ad content, with \texttt{plannedparenthood.org} (an American nonprofit that provides reproductive and sexual healthcare and sexual education in the U.S. and globally) exhibiting a frequency (0.161\%), more than double that of the second most popular domain (\texttt{lifechoicesinc.org}; 0.063\%). 
The remaining top advertisers included 15 CPCs, 5 ACs, 6 pro-choice related sites, and 2 anti-choice related sites.  

Whereas the top advertising AC provider (\texttt{planned- parenthood.org}) is an organization spanning the whole country, the CPCs at the top of this list are often specific to a single location. 
For instance, \texttt{lifechoicesinc.org} lists its location in Conway, AR, \texttt{anchorofhopewi.org} in Sheboygan, WI and \texttt{lifeclinic.org} lists 3 locations in Michigan state. 

\begin{table}[t]
\caption{Top 30 domains in the dataset}
\begin{center}
\begin{tabular}{llr}
\toprule
Domain & Kind & \% ads \\
\midrule
plannedparenthood.org & ac & 0.161 \\
lifechoicesinc.org & cpc & 0.063 \\
states.guttmacher.org & pro-choice advocacy & 0.063 \\
anchorofhopewi.org & cpc & 0.047 \\
lifeclinic.org & cpc & 0.040 \\
reproductiverights.org & pro-choice advocacy & 0.038 \\
prestonwoodpregnancy.org & cpc & 0.031 \\
noisefornow.org & pro-choice advocacy & 0.023 \\
guttmacher.org & pro-choice advocacy & 0.020 \\
alphacentermychoice.org & cpc & 0.019 \\
fpawomenshealth.com & ac & 0.017 \\
info.carafem.org & ac & 0.016 \\
hesperian.org & pro-choice advocacy & 0.014 \\
womenscarecenter.org & cpc & 0.013 \\
nyc.gov & government resource & 0.013 \\
wholewomanshealth.com & ac & 0.012 \\
abortionclinicservicesatlantaga.com & ac & 0.011 \\
care.wholewomanshealth.com & ac & 0.010 \\
abortiontestimonials.com & anti-choice advocacy & 0.010 \\
freeabortionalternatives.com & cpc & 0.010 \\
lifeforwardcincy.org & cpc & 0.009 \\
scprc.com & cpc & 0.009 \\
lifechoicesmemphis.org & cpc & 0.009 \\
carenetfrederick.org & cpc & 0.009 \\
arkadelphiapregnancy.com & cpc & 0.009 \\
alternativespc.org & cpc & 0.008 \\
womenonweb.org & pro-choice link to care & 0.008 \\
youroptions.com & anti-choice advocacy & 0.007 \\
go.fcws.org & cpc & 0.007 \\
go.option1.org & cpc & 0.007 \\
\bottomrule
\end{tabular}
\label{tab:top_domains}
\end{center}
\end{table}

\subsection{Relation to Laws}


\begin{figure}[t]
\centering
\includegraphics[width=0.79\columnwidth]{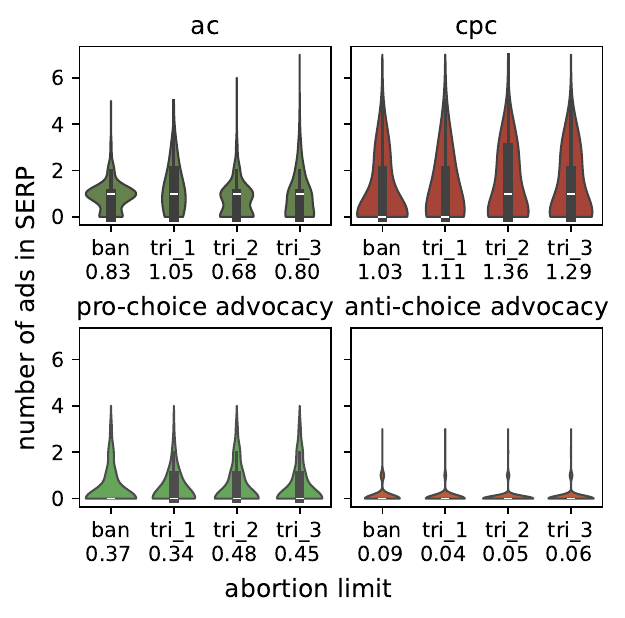}
\caption{Distribution of ad counts per SERP (y-axis) for four of the main domain type labels (see plot titles) in states with different abortion limits (x-axis). The mean of each distribution is shown below each x-axis tick label.}
\label{fig:ads_by_laws}
\end{figure}

Next, we examined the search results with respect to each state's abortion policy environment.
In Figure \ref{fig:ads_by_laws} we compare the distribution of different ad types separately for each policy environment, separated into ban, or limit at first, second, or third trimester. 
We examined the four most prevalent categories of ads in our dataset: ACs, CPCs, pro-choice, and anti-choice advocacy/research. 
On average, the number of AC ads was highest in states with first trimester ban, whereas the number CPC ads generally grew with increasing trimester limits. 
Given the large corpus of data, tests of difference were all highly significant ($p<0.01$), but the magnitudes of differences were fairly modest. 
The distribution for each ad type was similar across policy environments, regardless of gestational limits.
Notably, and not visible in this plot, we found that states with total bans had a higher proportion of AC and pro-choice advocacy advertisements than states with first trimester gestational limits.

\begin{figure}[t]
\centering
\includegraphics[width=0.97\columnwidth]{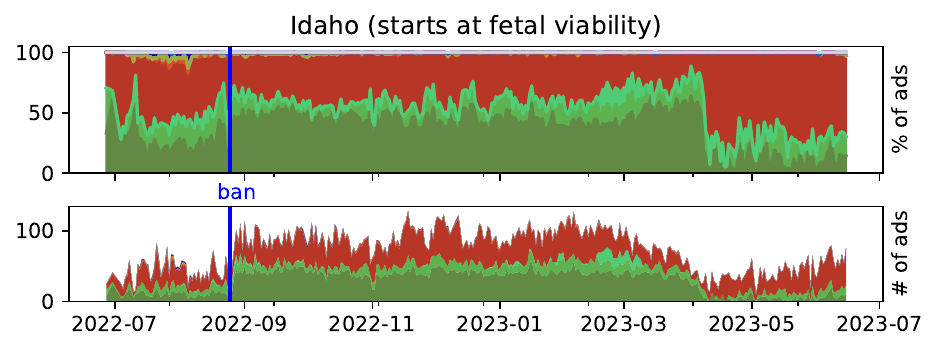}
\includegraphics[width=0.97\columnwidth]{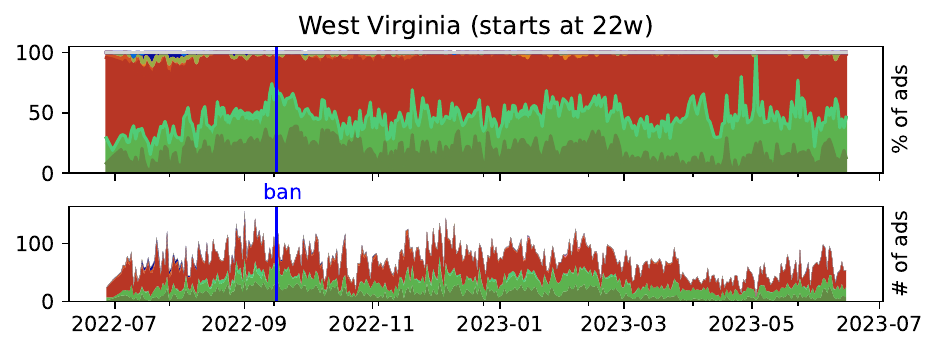}
\includegraphics[width=0.97\columnwidth]{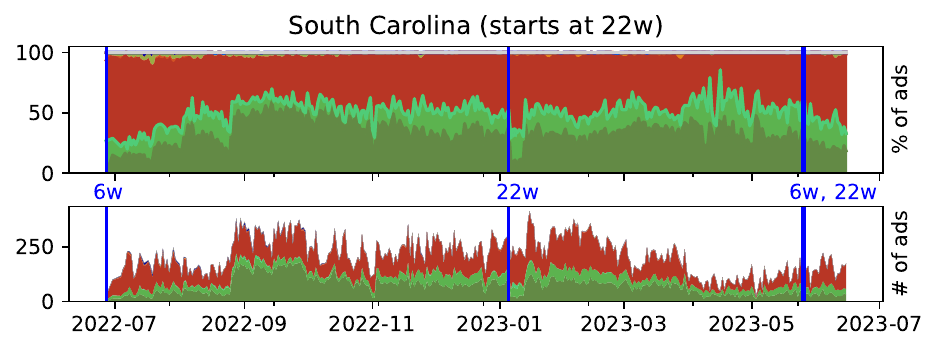}
\includegraphics[width=0.97\columnwidth]{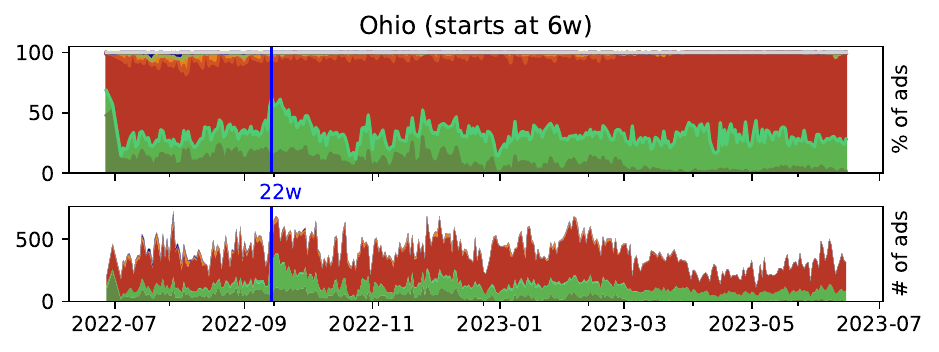}
\caption{Volume of ads returned (bottom) and proportion of ads (top) by kind for select states. Colors as in previous figures. Changes in laws indicated by vertical blue lines and text indicates the new abortion term limit.}
\label{fig:domain_kind_by_time_and_state}
\end{figure}

We then explored the relationship between the types of ads returned for our abortion-related queries and the states in which the searches were executed.
Figure \ref{fig:domain_kind_by_time_and_state} shows the total and proportional volume for the four treatment states used in the SASCM: Idaho, West Virginia, South Carolina, and Ohio.
Note that the raw number of results (bottom panels) depends on the number of voting districts in each state, and thus these are not comparable between states. 
Instead, we direct the reader to examine the trends in the proportion of advertisement types within each state (top panels).
Plots for all states can be viewed in the online supplementary material.\footnote{\url{https://tinyurl.com/AbortionAdsByState}} 
In these plots, we also indicate the time at which a new policy takes effect (vertical blue lines, along with text specifying the new abortion term limit).

\begin{figure}[t]
\includegraphics[width=0.95\columnwidth, right]{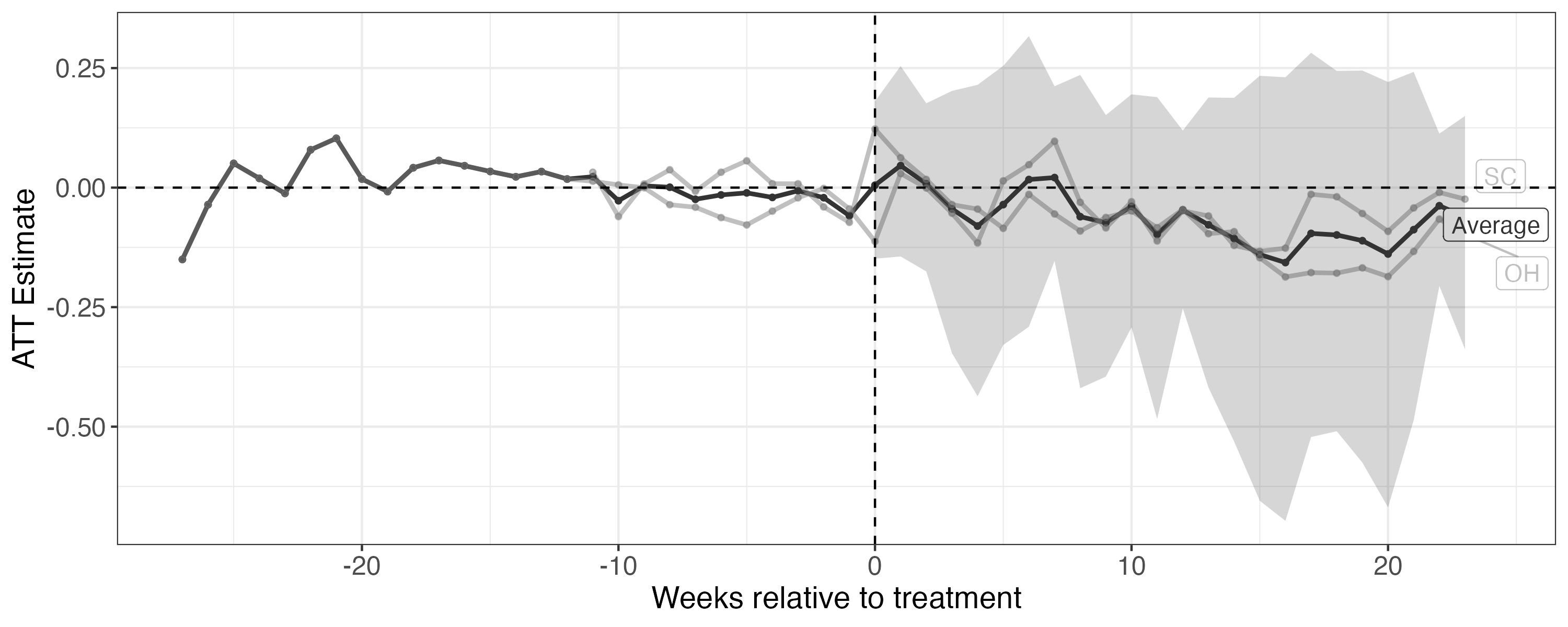}\\
\includegraphics[width=0.93\columnwidth, right]{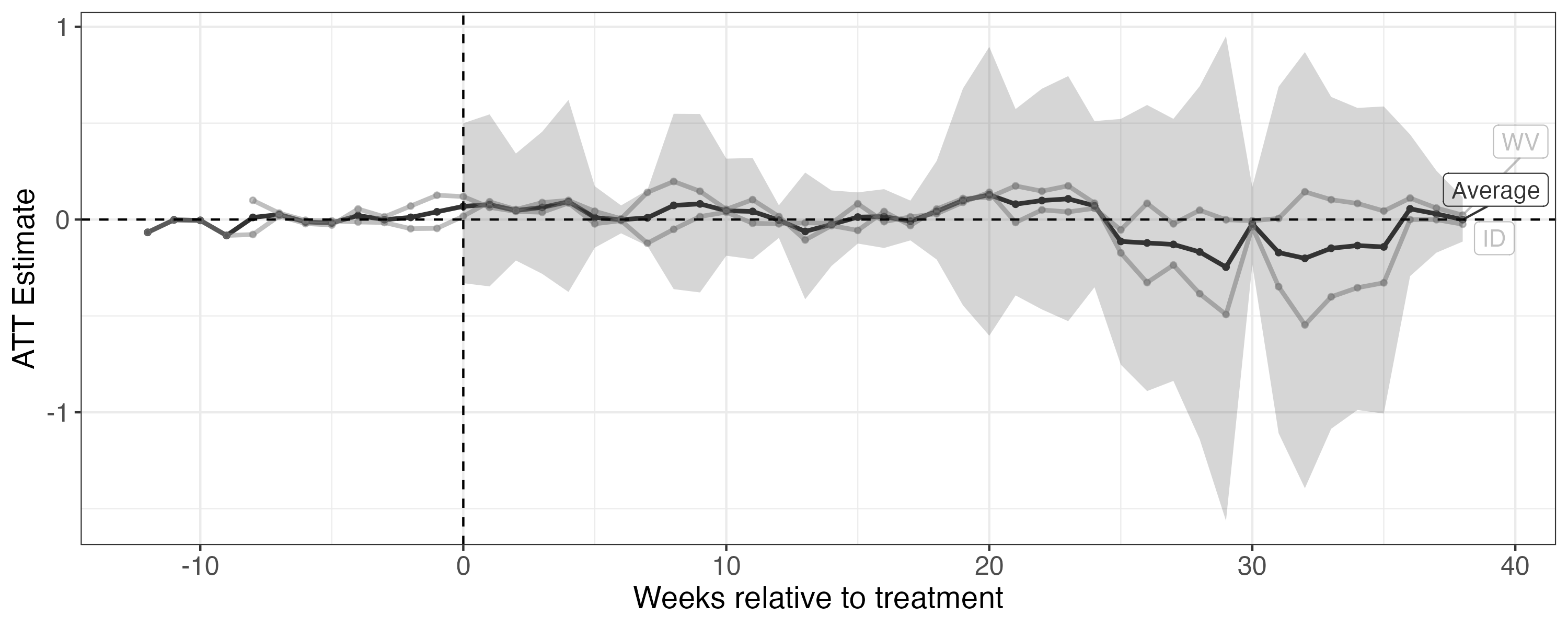}
\caption{SASCM models measuring effect of law change on the proportion of pro-choice ads. Top: Relaxed model with SC and OH as treatment states. Bottom: Restricted model with WV and ID as treatment states.Note that the weekly aggregation of policies eliminated two brief policy changes in South Carolina that lasted less than the majority of the week. Vertical dashed lines indicate the policy change. Solid black lines represent weekly average ATT estimates over time, with 95\% confidence intervals in grey. Solid grey lines represent weekly ATT estimates over time per treatment state. An ATT of 0 is a null effect. }
\label{fig:sacm}
\end{figure}

Clear patterns in changes of ad prevalence in response to policy changes were not visibly evident. 
Of note, Idaho experienced a marked increase in the relative proportion of CPC ads toward the end of the study period, however this was not precipitated by an abortion policy change. 
Findings from SASCM models using binarized ad and policy variables aggregated to the weekly-level further confirmed these findings, with small effect sizes that did not reach statistical significance for the proportion of pro-choice ads (overall ATT for restrictive model $= -0.01$ [SE $= 0.06$], overall ATT for relaxed model $= -0.06$ [SE $= 0.14$]). Figure \ref{fig:sacm} shows plots of weekly ATT estimates for each of these models. 

\section{Discussion}
\label{sec:discussion}

This study assessed trends in ads returned by Google Search for abortion-related queries during the year following the Dobbs v. Jackson decision in the US. 
Our analysis revealed several key patterns in advertisement content. 
First, ad content was sensitive to query type, with informational and safety-related queries about abortion returning the highest proportion of ads related to CPCs and anti-choice organizations. 
Conversely, queries about abortion legality returned a higher proportion of content related to pro-choice organizations.
Second, we found a relatively consistent balance between pro-choice and anti-choice advertising content across all queries over time.  
Third, state-level analyses revealed geographic variation in ad content, however, we have not attempted to link it to particular demographic or sociopolitical characteristics. 
Notably, states with total abortion bans had a higher proportion of AC ads. 
Further the average number of AC ads was highest for states with first trimester gestational limits and the average number of CPC ads was positively correlated with increasing gestational limits, suggesting possible reactive advertising strategies. 
Despite these patterns, our statistical analyses found no significant relationship between state policy changes and shifts in ad content over time.

The dominance of CPC ads in response to informational and safety-related abortion queries raises concern about access to accurate medical information among those seeking care.
Users seeking basic information about abortion procedures, who may be most vulnerable to misinformation, would be disproportionately exposed to ads from organizations that do not provide abortion services and often actively discourage them \cite{montoya2022problems}. 
Access to accurate medical information varies significantly based on query type and location, potentially limiting access to time-sensitive care when users need it most \cite{gupta2023trends}. 
Moreover, the increased presence of CPC ads in areas with higher gestational limits suggests a targeting of regions where abortion services are more legally accessible, potentially aiming to intercept care-seeking individuals \cite{polcyn2020truth}.

These findings also highlight the role of digital advertising policies in shaping access to information seeking around abortion. 
Despite Google's patchy efforts to label abortion providers, which were rolled out in June 2022 \cite{bloomberg2022}, CPCs maintain a strong ad presence on the platform. 
The consistent presence of CPC ads alongside ads for medical services, particularly in response to searches about abortion information and safety, therefore may exacerbate confusion among users seeking medical information \cite{montoya2022problems}. 

Examination of state-level data revealed unexpected patterns in abortion-related ad content. 
Further, despite shifts in state abortion policies, we did not find significant reciprocal changes in the proportion pro-choice advertisements over time. 
However, some variations emerged when analyzing ad composition relative to different policy environments. 
Notably, states with total abortion bans showed a higher prevalence of AC ads, while states with later-term abortion limits saw a greater average number of CPC ads. 
This suggests that advertising strategies may be shaped more by market conditions than by immediate policy changes.
This may also indicate strategic advertising plans, such that pro-choice organizations prioritize outreach in restrictive environments to maintain awareness of services that may be available in neighboring states \cite{npr2022}.
Ultimately, we found that access to accurate reproductive health information via search engines is unevenly distributed across the US, potentially reinforcing existing disparities in abortion access \cite{Mejova_Gracyk_Robertson_2022}.

This study has many methodological strengths. 
First, it provides a unique analysis of abortion-related ad patterns across the US during the year following the Dobbs v. Jackson decision, spanning 435 congressional districts and including over 3.6 million ads. 
Second, our systematic data collection process captured daily advertising patterns across multiple query types, from general searches to specific questions about legality, safety, and cost. 
Third, our domain classification approach achieved high inter-rater reliability and expanded upon previous labeling schemes to capture nuanced distinctions between different types of abortion-related resources. 
Fourth, our year-long data collection period, combined with daily tracking of state policy changes, enabled analysis of advertising patterns across varying regulatory environments.

Several limitations should also be acknowledged when interpreting these results. 
Our data collection was confined to Google search results and did not account for information dissemination through other digital platforms, indicating a need for research examining advertising patterns across broader online spaces. 
Our query set, though systematically developed through Google's search features, may not have captured the variance in real users' queries. 
Furthermore, the requirements of SASCM models, specifically the need for adequate pre-treatment periods and comparable control states, limited our analysis of policy effects to a subset of states experiencing policy changes during the study period. 

Future work should evaluate the effectiveness of current advertisement labeling practices and their impact on information access. 
Studies incorporating real-world search patterns and user-focused investigations could provide insight into individual-level engagement with abortion-related advertising under ecologically-valid conditions \cite{feal2024introduction,greene2024current,robertson2023users}. 
Additional statistical approaches could examine associations between advertising patterns and state characteristics, including demographic, political, and healthcare access variables. 
Finally, while our analysis covered a substantial period following Dobbs, continued monitoring of online information ecosystems is needed to understand long-term shifts in abortion-related advertising strategies \cite{munger2019limited,munger2023temporal}.

Our findings demonstrate that Google search advertising patterns for abortion-related queries vary systematically by query type and location, with CPC advertisements appearing prominently in response to several queries. 
This observed prominence of CPC advertisements in search results may create barriers for users seeking time-sensitive abortion-related information and services and highlights the critical role that digital advertising policies play in shaping access to abortion care.
These advertising patterns raise questions about user access to accurate abortion-related information on Google Search and suggest a need for further examination of search engine advertising policies, specifically regarding the placement of advertisements from organizations that do not provide medical services. 

\section{Acknowledgments}
\label{sec:Acknowledgments}
We would like to acknowledge Kara E. Rudolph, PhD, Columbia University, 
for her 
consultation on statistical methods, and Andrew Schwartz
for his assistance in putting together the database of abortion-related policy changes following Dobbs v. Jackson.
The search data used in this study were collected using machines at Northeastern University 
that are administered by the authors' collaborators, with their permission. 
Northeastern University was given permission to query Google Search automatically for research purposes. Google did not review our research design, nor had any review rights with respect to the manuscript.
Jeffrey Hancock was the faculty director of this work at Stanford.

\bibliographystyle{IEEEtran}
\bibliography{references}

\end{document}